Article

# A survey of location inference techniques on Twitter

**Oluwaseun Ajao**
School of Electronics, Electrical Engineering & Computer Science
Queen's University Belfast, United Kingdom

**Jun Hong**
School of Electronics, Electrical Engineering & Computer Science
Queen's University Belfast, United Kingdom

**Weiru Liu**
School of Electronics, Electrical Engineering & Computer Science
Queen's University Belfast, United Kingdom

**Abstract**
The increasing popularity of the social networking service, Twitter, has made it more involved in day-to-day communications, strengthening social relationships and information dissemination. Conversations on Twitter are now being explored as indicators within early warning systems to alert of imminent natural disasters such earthquakes and aid prompt emergency responses to crime. Producers are privileged to have limitless access to market perception from consumer comments on social media and microblogs. Targeted advertising can be made more effective based on user profile information such as demography, interests and location. While these applications have proven beneficial, the ability to effectively infer the location of Twitter users has even more immense value. However, accurately identifying where a message originated from or author's location remains a challenge thus essentially driving research in that regard. In this paper, we survey a range of techniques applied to infer the location of Twitter users from inception to state-of-the-art. We find significant improvements over time in the granularity levels and better accuracy with results driven by refinements to algorithms and inclusion of more spatial features.

**Keywords**
Location Inference; Twitter analytics, Information retrieval

## 1. Introduction

The ability to accurately profile the location of social media users comes with immense benefits to service providers and consumers themselves [1]. This continues to be a well-explored research domain [2]. The dilemma of correctly identifying the author's location combined with the unique language of microblogs such as Twitter, Facebook and Foursquare has brought with it some challenges that were not associated with structured texts and online blogs, forums and conventional online media.

Twitter now has more than 300 million monthly active users who on a daily basis generate over 500 million conversations popularly referred to as 'tweets'[1] 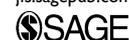 which are text messages consisting of a maximum of 140 characters. The limited space requires brevity in writing, giving rise to an informal dictionary of words only used within the social media space. In addition, writing on Twitter tends to have lots of non-standard abbreviations, typographical errors, use of emoticons, irony, sarcasms and trending topics referred to as *hashtags*. Such unconventional, unstructured texts are regarded as noise as standard natural language processing (NLP) tools do not handle such well [3], leading to an interesting challenge in tweet content analysis.

**Corresponding author:**
Oluwaseun Ajao, School of Electronics, Electrical Engineering & Computer Science, Queen's University Belfast, BT7 1NN United Kingdom
Email: oajao01@qub.ac.uk

Location inference on Twitter can be used to identify offenders engaging in bullying of other online users via social media also known as 'cyberbullying'. Also Twitter has been known to be a good platform for detecting the outbreaks of diseases and natural disasters. The ability to accurately infer the location of affected users can save lives and help in crisis management.

It has been shown in [4] that Twitter serves as a platform for building social relationships and is utilitarian for information purposes. In the former, users having a bidirectional *following* relationship (allowing the follower's public posts to be continually displayed on the follower's news feed) called 'friends'. In the latter, a unidirectional following relationship exists where a user may only follow another influential user they are interested in. However, the followee may choose not to reciprocate the gesture thus being a unidirectional relationship. This is more common with corporations, celebrities, public figures and politicians who may have a significantly larger number of followers and just a handful of friends.

Twitter users have the option to disclose their city level location which should normally be their primary residence. Text may be input within a location field as part of their Twitter user account registration. In reality, less than 14% of users accurately complete this field. In [5], it is discovered that 34% of Twitter users gave false or fictitious location names. Because this is an optional, free text field, Twitter does not regulate or enforce what their users can input.

Also, to enhance the experience of its users, Twitter allows inclusion of location coordinates as metadata to tweets. This is called geotagging; the current location of the user can be included in tweets sent from mobile devices. Geotagged messages can give an accurate estimate of the current location of the user or the origin of a particular tweet up to the nearest kilometre. Similarly, even though virtually all recently manufactured smartphones now come with a GPS, less than 0.5% of Twitter users turn on the location function of their smartphones due to concerns over privacy [6], cyber bullying and stalking. Other users switch off location services to conserve power and prevent their batteries from running out quickly [7].

Various works have employed diverse kinds of spatial features to infer the location of online users including use of metadata information such as time of post [8]. Some have used only the content of the tweets [9][10][11]. The others have looked at the social network relationship amongst users [12]. The user account information has also given useful insights for this purpose [13][14], while some have followed a hybrid approach [15]. There is also a growing trend for the use of location-based social networks [16][17][18]. However most works observed still tend to include the message text as a key input for their study and techniques.

Techniques have ranged from natural language processing including named entity recognition (NER), parts of speech tagging (POS) [19], machine learning and probabilistic methods [20] as well as gazetteers and location databases [21]. Results achieved by the various works are diverse and have been shown to be getting higher granularity levels with average error distances of less than 1 km [16]. To the best of our knowledge, our work is the first to survey the various techniques employed in this field.

The rest of this survey is organised as follows. In Section 2 we will discuss the types of locations on twitter. Spatial indicators and location features that have been used are discussed in Section 3. Section 4 further reviews the types of techniques employed in location inference while Section 5 looks at the data collection and analysis process. Conclusions and future work are presented in Section 6.

## 2. Types of location on Twitter

Initial works in the field of location inference made no differentiation between the home residence of a Twitter user and their current location. It is observed that some authors had earlier referred to it as '*User Location*' [12][9][22] assuming the geotagged location of the tweet to be the user location. [5] infers the home residence of the user to be already contained within the location field provided as part of the user account information. [23] cites the fact that it was possible to tweet about a particular location and not be in that location at the time. It illustrates the concept of space and time. [16] examines the concept of determining points of interests (POI) with a temporal awareness of the past, present and future as mentioned in the message text.

[23] also defines 4 distinct location types on Twitter, namely, locations directly mentioned in the message text, focused locations i.e. described by the message context, user's current location (from where a tweet was sent) and their location profile which can be a combination of their current, previous home locations and other places they frequently visit . A diagrammatic illustration of locations inferred on Twitter is given in Figure 1.

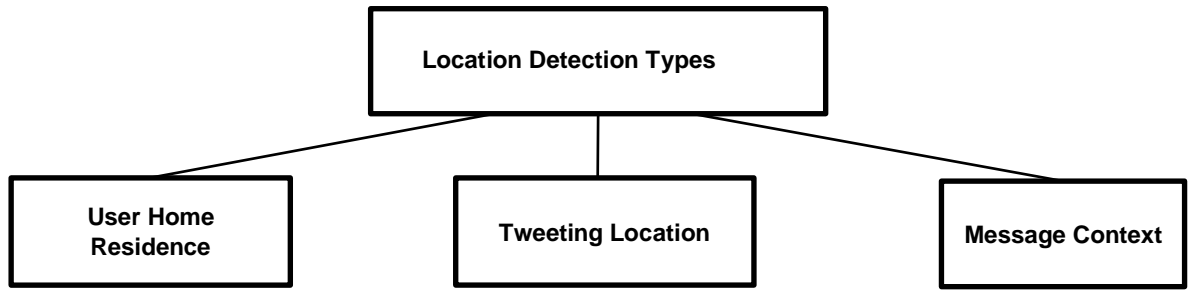

**Figure 1**.Types of locations inferred on Twitter

## 3. Spatial features and indicators

As illustrated in Figure 2, diverse indicator types that help to infer the location of Twitter users have been employed over the years and we shall look at them in more detail.

### 3.1. Message Context

Twitter message text forms the backbone of most research within the field of location inference as this helps understand the context of the messages themselves. The challenges associated with tweet text processing can be very much linked to the unstructured format of those messages as opposed to online articles and blogs that have more content and follow conventional grammatical and semantic usage. These include abbreviations and more so non-standard ones as there is no precise rule of writing on the social media platform. Because most of tweets are sent via mobile devices their users have a large leeway for typos and brevity. An instance would be the abbreviation for the United Kingdom which could be *'UK', 'GB', 'GBR' or 'GR8 Britain'*. [16] uses the Brown clustering to handle out-of-vocabulary (OOV) words. While [11] uses the Jaccard coefficient to resolve and accommodate similar words. [18] uses cosine similarity to match actual location with a list of keywords. A good content analysis approach would take into consideration all possible instances of the location entities being expressed within the message. It is important to note that even when locations are identified within the messages, it cannot be automatically inferred as the user location or even the tweet location [18]. A good example would be where a tweet contained the city name 'Belfast'; however it may not necessarily imply the author was based in Belfast or that the tweet was even sent from Belfast.

   Some works have used the URL links within the body of the text as spatial indicators for inferring the location of the users. [17] uses these links to infer the country level location by inputting the corresponding domain server IP addresses into the InfoDB database - a free online query service that matches geographical location with IP addresses and domain names. The most successful techniques have employed use of the message content alongside one or two other features to have a robust output.

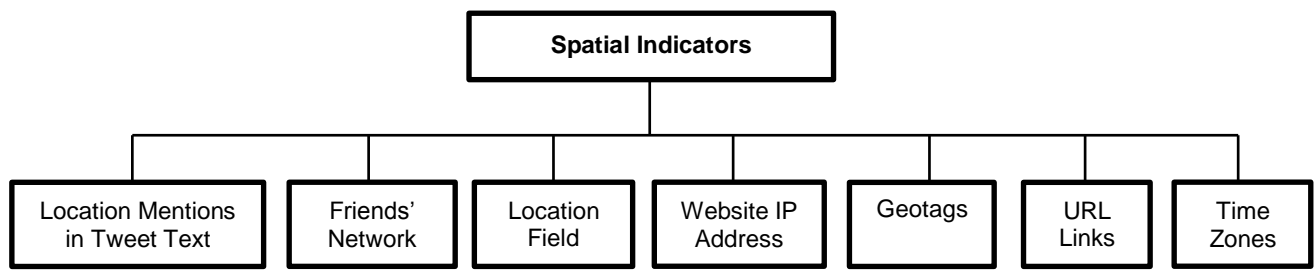

**Figure 2.** Indicators of user location

## 3.2. Social Networks

The followers of a user have been shown to be a good indicator of their home residence. While reciprocal following relationship can provide evidence of strong user connections, other indicators can include regular exchange of messages or frequent mention of each other's names within messages. [15][24][4] have shown that two users are likely to communicate frequently if they reside within the same city and vice versa.

[6] mentions the possibility of having multiple location profiles based on the user's offline social relationships with other users. According to [20] the more influential a user is, the higher the diversity of their followers and friends would be from around the world. [12] shows that the network of a user would be optimal for inferring location up to the third depth.

## 3.3. User Profiles

The account information given at the point of registering a Twitter user account can give very useful insight into their location allowing advertisers to accurately target their customers. It can also help emergency services and first responders to immediately locate the scene of a crisis or disaster or to help track down potential offenders in cyber bullying crimes. Usually the location field follows a free text format enabling the users to manually type in their city name. It would normally be in the *City or State* granularity level such as Glasgow, Scotland. However, instances of less conventional phrases such as the 'The Big Apple' or even meaningless expressions such as 'Bieber Town' make it difficult for conventional NLP and machine learning algorithms to effectively extract the location entities and in some instances are likely to give misleading results.

The user's website or personal web page could also be listed on the account information and would normally hold useful information. This would be so in particular if the website listed by the user was hosted by a provider resident within their home country and with possibly city-level information, if they resided in the same city. However, there is the possibility of hosting their website in one geographical location and living elsewhere. For instance a user based in the US might had initially signed up for web hosting with a provider based in the US but if they relocated to say, Australia but had not switched service providers. This would mean that their web domain and server IP address would still be indexed to their former country of residence which is the United States whereas they currently reside in Australia.

## 3.4. Geotags

Most smartphones are now equipped with the global positioning system (GPS) function as a standard feature and working with this, geo-satellites are able to accurately pinpoint the user's geographical location i.e. latitudes and longitudes coordinates. This would usually be an optional feature for users to enable due to their privacy concern and it has been found that less than 0.5% turn on their location services [20] making this a challenging feat. This indicator is very useful where the user is mobile and frequently updates their location profile. [15] uses Vincenty's geometric median an estimate well applied to the field of Geography and land surveying [25] to estimate the location of a Twitter user using their last 5 geotagged tweets that occurred within a 15km radius, as shown in (1).

$$m = \text{argmin}_{x \in L} \sum_{y \in L} Distance(x, y) \qquad (1)$$

## 3.5. Third Party Sources

The popularity of location-based social media sites has enabled means of interaction also referred to as Geosocial networking. Foursquare and Yelp are good examples of these sites offering companies, small businesses and restaurants the opportunity of registering on their directory which gets such businesses enlisted as part of a geographic database. Online users are able to find the location of a place of interest, say a restaurant in Belfast, simply by searching their online directory. Previous visitors to these locations are able to leave reviews and comments about these places called 'check-ins'. Foursquare allows its users to connect their Twitter accounts to Foursquare posts which are usually geotagged thus allowing to infer their location from a Foursquare message post even though they have not disclosed their location on Twitter [16].

## 3.6. Time Zones

Tweets metadata usually contain a timestamp of the message and the time zone as captured by the Twitter API. This is a useful feature that can allow the inference of the location to at least country-level granularity [2]. This would be quite useful where there is limited and sparse location information within the body of the message text.

## 3.7. Web Snippets

[8] addresses the sparsity problem of tweets in locating points of interests by employing webpage snippets. Rae, [26] searches Wikipedia to get structured information about places to complement tweets about points of interests (PoIs).

# 4. Methods of inferring locations on Twitter

Diverse approaches and techniques have been used in the past and are currently being employed to better improve the accuracy of location inference methodologies and algorithms. This burgeoning field lends techniques ranging from several fields of study involving machine learning, statistics, probability, natural language processing to geographical information systems and surveying. Diverse methods have achieved varying levels of success; in any case the effectiveness and granularity levels achieved by these methods continue to improve rapidly.

However, the informal nature of the social media platform as well as unique language of expression brings with it some challenges in trying to properly deduce the meaning and context of these conversations. They contain frequent use of emoticons, sarcasms, hashtags, abbreviations and typographical errors. This leads to the need for robust methods and algorithms that will factor that into its input.

In the analysis of text messages, names of places mentioned could be ambiguous. For example the word 'Washington' could refer to the state or a place bearing the same name within the District of Columbia both in the United States. Washington DC and Washington State are 3,000 miles apart. The process of trying to disambiguate place names is called 'toponym resolution'. It becomes more complicated when noun types could have similar names; for example, a person could also be called Washington. Techniques used in location inference can be broadly grouped into three categories namely natural language processing, machine learning and use of location databases or gazetteers.

## 4.1. Natural Language Processing (NLP) Techniques

Natural processing methods applied include the named entity recognition which could be either segment-based or word-based representation [16] with the former showing more effectiveness in recognising entities within tweets and the widely used tool for this technique is the StanfordNER. [27] found that use of the StanfordNER on social media texts did not accurately detect entities including location names, especially if they were unusually abbreviated thus having a high probability of type I error (false negatives). However, [19] retrained four NER tools namely StanfordNER, OpenNLP, TwitterNLP and Yahoo! Placemaker on 2,878 disaster-related tweets applying a 10-fold cross validation and found the retrained StanfordNER to have the highest F-measure of 0.9. In [39] a hybrid approach was adopted where location entities were extracted and parsed into a gazetteer to accurately geocode the place names mentioned in the tweets.

The conditional random field (CRF) technique is recommended for handling complex dependencies within phrases and sentences. The University of Illinois at Urbana Champaign NETagger [28] has also been well used till date. NLP techniques often tend to be applied with probabilistic tools such as multinomial Bayesian and generative probability models. It requires training data and may be complex to apply. However, it allows the development of sophisticated algorithms that suit the user's needs. It has also been shown to have a quicker processing time. Another benefit of using NLP is its flexibility in identifying unconventional words (which is quite common on Twitter) as similarity checks between words can be done in order to identify entities listed within a keyword list [18].

## 4.2. Gazetteers

Gazetteers and Geographical Databases are also well applied to the study and some tools used include the United States board on geographic names popularly called GeoNames[2], GeoNet[3] and the US census TIGER Gazetteers[4]. Some works have also used a hybrid of the earlier mentioned techniques. For example [29] proposed a system for inferring the current location of a Twitter user using the PipePOS tagger and the USGS location database to resolve ambiguous location names.

Gazetteers are easy to implement [17]. Also they do not require training data but there is a challenge of slow processing speed. [12] shows that varying the size of their Twitter dataset by increasing the depth of friends/followers relationship has no impact on the time taken to compute and detect the location of a tweeter using the gazetteer method. This can be especially frustrating in databases with very large dataset. Thus there also exists the challenge of toponym resolution and matching of words with the location database to cater for the abbreviations and unconventional writing style on Twitter and in most cases location names which are found in messages but do not exactly match the database thus are discarded and could lead to a type I error (false negative).

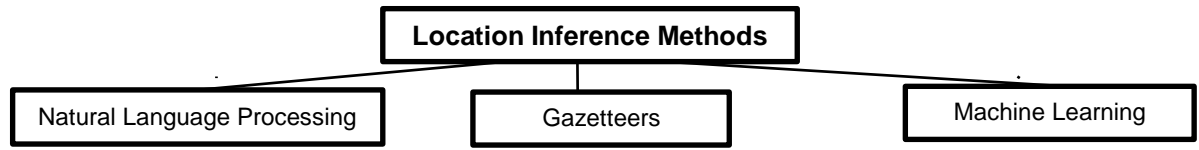

**Figure 3.** Main categories of location inference techniques

### 4.3. Probabilistic and Machine Learning Techniques

Techniques for the detection of location of Twitter users have also been adopted from data mining and machine learning techniques. It has been shown to be a good method of clustering Twitter users [30], using *k*-nearest neighbour, fuzzy matching [12], Naives Bayes, probabilistic clusters, Markov chain models etc. [31] used a probabilistic model that incorporated the local words used by users while users who had not mentioned sufficient local words had their location inferred from the local words of their friends network. Also, in [9] [10] [11] location was inferred from probabilistic distribution of users local words.

[13][20] use a probabilistic algorithm based on friends relationship. [15] uses a graph-based approach applying label propagation to predict location from that of other users in their network. [32] develop geographic and t*opic* models adopting Mean Field Variational inference and Kullback-Leiber divergence. [5] proposes a Naives Bayes model classifier. [18] learns the patterns of location based services from past messages to predict current location. [1] developed language models using Bayesian inversion. [16] used a CRF classifier to identify points of interest (PoIs) incorporating four classes i.e. lexical, grammatical, geographical and BILOU schema features. In [6] a model that considered both the tweeting and following relationships was used. [28] looked at dynamically weighted ensemble method to create a combination of Naives Bayes, Naives Bayes Multinomial and Heuristic classifiers that can predict user location at all levels of granularity.

## 5. Tweet Gathering and Analysis

Tweets made public are usually accessible in the online domain method and can be retrieved using the Twitter REST API[5] while live updates on individual or multiple users can be extracted as required in real-time using its streaming API. This accessibility makes Twitter a powerful tool in the gathering and analysis of public views allowing its users to become social sensors within the population.

### 5.1. Tweet Corpuses

Corpus sizes of tweets gathered have varied from relatively small datasets of under 62,000 tweets [5] to as large as 615 million tweets [31]. Time span of the data collected was usually in the range of few weeks to a couple of months. On the one hand, the REST API is also useful for the collection of specific user tweets allowing for the backtracking of their timeline to gather their most recent 3,200 tweets. At the time of writing this paper, the Twitter search API allows the collection of tweets by defined keywords or around a specified location name or coordinates (geotagged messages) for tweets posted up to the previous 6-9 days. On the other hand, the streaming API that collects the messages as they are being broadcast would only be able to receive 1% of the *Firehose*. Twitter data partners such as GNIP[6] or Datasift[7] provide a premium service that supplies messages covering a longer duration as well as 100% access to the *Firehose*.

**Table 1.** Datasets and collection periods of some works.

| Reference | Corpus Size | Period Covered | Duration (Months) |
|---|---|---|---|
| [14] | 2,495,000 | Jan '11 – May '11 | 5 |
| [32] | 380,000 | Mar '10 | 1 |
| [5] | 62,000 | Apr '10 – May '10 | 2 |
| [15] | 47,700,000 | Apr '12 – Nov '12 | 8 |
| [16] | 4,330,000 | Jun '10 | 1 |
| [33] | 1,524,000 | Jul '11 – Aug '11 | 2 |
| [4] | 100,000,000 | Jun '10 | 1 |
| [23] | 20,000,000 | Apr '11 | 1 |
| [31] | 615,000,000 | Jun '10 – Apr '11 | 13 |
| [17] | 80,000,000 | Sep '11 – Feb '12 | 6 |

Another means of gathering Twitter data for training and testing location inference algorithms would be from other researchers within the field. An example is the Social Network Analytics Platform[8] (SNAP) provided via open access by Stanford University. It includes large tweet corpuses and social networking data which can be used for graph analysis.

### 5.2. Results and Metrics

The results achieved by various works have significantly improved over time with regards to increased accuracy and granularity levels. This has been largely driven by refinements to algorithms and inclusion of more spatial features. In the same vein, the effectiveness of spatial features and/or accuracy of the algorithms required to achieve finer granularity levels increase progressively for time zones, country, region, city and post codes respectively. For example a more accurate prediction method would be required to estimate a Twitter user's home postal code as opposed to one that infers their country of residence.

Several metrics have been presented to compare the performance and results of the methods with one another. They include accuracy within a specified range say 10 km, error distance, average error distance (AED) and median error distance (MED). To validate the effectiveness of the methods against other baselines, the *k*-fold cross validation has been well utilized while the precision, recall and F-measure i.e. harmonic mean of both indices are derived. Table 2 shows the growing trend of finer grained location inference on Twitter.

Over time, accuracy levels and granularity of results have continued to improve starting from 2010 when inference was only precise to the city-level. This resulted from the fact that location was inferred solely on the basis of the tweet content without giving consideration to other information such as web links, friend the user profile and other metadata associated with the message, however with the subsequent adoption of spatial features such as user check-ins gathered from location-based services including Foursquare, accuracy has improved significantly with the most recent work published four years later [31] achieving a 60% accuracy within a 10km. This is a remarkable improvement as opposed to a performance of 51% accuracy over a 160km radius recorded by [11].

**Table 2.** Improvement in granularity levels over the past 5 years.

| Reference | Year | Technique | Accuracy (%) | Coverage radius | Location Type |
|---|---|---|---|---|---|
| [11] | 2010 | Probabilistic (ML) | 51.00 | 160km | User location |
| [32] | 2010 | Geographic topic model (NLP) | 24.00 | State level | Home location |
| [1] | 2011 | Language models | 13.90 | Zip code level | Tweet location |
| [1] | 2011 | Language models | 29.80 | Town level | Tweet location |
| [10] | 2012 | Gaussian Mixture models & Maximum Likelihood Estimation (ML) | 49.90 | 160km | Home location |
| [20] | 2012 | Probabilistic (ML) | 62.30 | 160km | Home location |
| [18] | 2012 | Machine learning | 20.00 | 10km | Tweet location |
| [34] | 2012 | Dynamic Bayesian Networks | 57.00 | 0.1km | Home location |
| [17] | 2013 | Gazetteer | 37.00 | 10km | Tweet location |
| [31] | 2014 | Probabilistic (ML) | 60.00 | 10km | Users main location |

$$\text{Precision} = \frac{\text{True Positives}}{(\text{True Positives} + \text{False Positives})} \quad (2)$$

$$\text{Recall} = \frac{\text{True Positives}}{(\text{True Positives} + \text{False Negatives})} \quad (3)$$

$$\text{F - Measure} = \frac{2 * \text{Precision} * \text{Recall}}{(\text{Precision} + \text{Recall})} \quad (4)$$

## 6. Conclusion and Future Work

Location inference can be applied to many areas and its applications include marketing and consumer user profiling. The importance and popularity of location-based social networking services continues to grow as billions of videos are being uploaded daily and shared worldwide on Twitter and other social networking platforms. [35] extended the Hyperlink-Induced Topic Search (HITS) algorithm to identify and rank the relationships existing between a set of keywords (tags) and a set of location-aware content such as videos and photos on Flickr. This further illustrates the need to accurately *map* topics and conversations to related location resources within the broader social media space.

[36] applied semantic information gathered from tweets to develop a system that detects and provides early warning alerting its users of an earthquake occurring in a location. The location accuracy of such a system is crucial for first responders and for emergency medical services to formulate effective evacuation strategies. Streamlining the detection in these locations would mean a more efficient and effective earthquake detection system.

The early detection of an epidemic outbreak is hinged upon a surveillance system that effectively captures the prevalence of syndromic conditions expressed by a population of interest. [37] shows there exists a positive relationship between tweet mentions of disease symptoms and public health data. Syndromic data gathered from tweets would be of immense benefit if spotted on time as an interesting pattern or anomaly and better still, if the precision of the location or the part(s) within the entire population is known or accurately inferred. Thus, Twitter user locations inferred and known on time could help forestall the spread of a deadly disease outbreak thereby saving lives. It will also ultimately save money as it would cost less to administer and treat infected patients if the disease is contained in its early stages of manifestation.

There are increasing reports of stalking and 'cyberbullying' where people are being verbally assaulted and at times sexually harassed by people they may or may not know. In most cases the users would veil themselves with anonymous user accounts with the belief that they cannot be identified. This continues to remain a challenge for police and law enforcement, proving to be even more difficult to produce sufficient evidence to prosecute such offenders in the court of law thus even more sophisticated technological methods such as cryptography are being applied [38]. There are cases that have led to the eventual suicide of their victims as well as the demise of offenders themselves [39]. This has prompted a lot of privacy concerns and raises questions as to how safe online social communication is. Also, potential applications of this would be better public enlightenment as to what level of information they should disclose online if they want to remain anonymous because their location could be implicitly inferred from other means such as content of their tweet messages, relationship with other users and their account information just to mention a few.

While some Twitter users would like to switch on the location services of their smart phones, there is the limitation of mobile device battery life thus some only enable the GPS function once in a while. However in event of a natural disaster such as an earthquake or a Tsunami, Twitter users may switch on this service [40] to support emergency rescue efforts. TEDAS is a system developed in [41] for the identification of crime and disaster related event (CDE) tweets while extracting the location from such messages from the user's past tweets as well as their friend networks using a rule-based classifier. It is expected that future work would look at ways of further improving the granularity levels of locations inferred on Twitter. Better algorithms would imply fewer friend network and information are then required to infer locations accurately.

## Notes

1. https://about.twitter.com/company
2. http://geonames.usgs.gov
3. http://www.geonet.org.nz
4. https://www.census.gov/geo/maps-data/data/gazetteer.html
5. http://dev.twitter.com/overview/documentation
6. http://www.gnip.com

7. http://www.datasift.com
8. http://snap.stanford.edu/snap.

**Funding**

This research received no specific grant from any funding agency in the, public, commercial or not-for-profit sectors.